\documentclass[letter]{aa} 

\usepackage{graphicx}
\usepackage{txfonts}
\usepackage{natbib}
\usepackage{amssymb}
\usepackage{multirow}
\bibpunct{(}{)}{;}{a}{}{,} 
\usepackage{pdflscape}
\usepackage{supertabular}

\begin{document}

   \title{Compact radio cores in radio-quiet AGNs}

   \author{A. Maini \inst{1,2,3,4}
          \and I. Prandoni \inst{3}
          \and R. P. Norris \inst{4,6}
          \and G. Giovannini \inst{1,3}
          \and L. R. Spitler \inst{2,5}
          }

   \institute{Dipartimento di Fisica e Astronomia, Universit\`a di Bologna, viale B. Pichat 6/2, 40127 Bologna, Italy
             \and Department of Physics and Astronomy, Macquarie University, Balaclava Road, North Ryde, NSW, 2109, Australia
             \and INAF-IRA, via P. Gobetti 101, 40129 Bologna, Italy
             \and CSIRO Astronomy \& Space Science, PO Box 76, Epping, NSW 1710, Australia
             \and Australian Astronomical Observatories, PO Box 915, North Ryde, NSW 1670, Australia
             \and Western Sydney University, Locked Bag 1797, Penrith South, NSW 1797, Australia
             }

   \date{Received 13/02/2016; accepted 18/03/2016}

   \abstract
    {The mechanism of radio emission in radio-quiet (RQ) active galactic nuclei (AGN) is still debated and might arise from the central AGN, from star formation activity in the host, or from either of these sources. A direct detection of compact and bright radio cores embedded in sources that are classified as RQ can unambiguously determine whether a central AGN significantly contributes to the radio emission.}
    {We search for compact, high-surface-brightness radio cores in RQ AGNs that are caused unambiguously by AGN activity.}
    {We used the Australian Long Baseline Array to search for compact radio cores in four RQ AGNs located in the Extended {\it Chandra} Deep Field South (ECDFS). We also targeted four radio-loud (RL) AGNs as a control sample.}
    {We detected compact and bright radio cores in two AGNs that are classified as RQ and in one that is classified as RL. Two RL AGNs were not imaged because the quality of the observations was too poor.}
    {We report on a first direct evidence of radio cores in RQ AGNs at cosmological redshifts. Our detections show that some of the sources that are classified as RQ contain an active AGN that can contribute significantly ($\sim$\,50\% or more) to the total radio emission.}

   \keywords{galaxies: active --
             galaxies: high-redshift --
             radio continuum: galaxies
             }

   \maketitle

\vspace{-0.3cm}

\section{Introduction} \label{sec:Intro}

Multi-wavelength studies of deep radio fields show that the sub-mJy population has a composite nature. While star-forming galaxies (SFG) dominate at $\mu$Jy levels (e.g.\ \citealt{Sey2008}), RL AGNs are the dominant component at flux densities >\,0.5\,mJy (e.g.\ \citealt{Mig2008}).

Recently, RQ AGNs have been shown to partly account for the flattening of the 1.4\,GHz source counts at flux densities <\,0.5\,mJy. RQ AGNs show AGN activity at non-radio wavelengths, but the origin of their radio emission is debated \citep{Bon2013, Smo2015}. 

Most RQ AGNs are unresolved or barely resolved at a few arcsec scale, which is similar to the host galaxy size. RQ AGNs have also been found to share properties with SFGs. They have similar radio luminosities (10$^{22-24}$\,W\,Hz$^{-1}$) and similar optical- or infrared-to-radio flux ratios. Their radio luminosity functions show similar evolutionary trends \citep{Pad2011a}, and their host galaxies have similar colours, optical morphologies, and stellar masses \citep{Bon2013}. For all these reasons it was concluded that the radio emission in such RQ AGNs is mainly triggered by star formation \citep{Pad2011a, Bon2013}.

Further support to this scenario was provided by \citet{Bon2015}, who found that in RQ AGNs radio-derived star formation rates (SFR) are consistent with FIR Herschel-derived SFR, though with a larger dispersion than for SFGs. This larger dispersion, on the other hand, does not rule out the possibility that some radio emission in these objects is associated with AGN activity.

It is well known that in the local Universe ($z$ $\lesssim$ 0.5) both AGN and SF processes can contribute to the total radio emission in RQ AGNs (e.g.\ Seyfert\,2 galaxies; \citealt{Roy1998}), and there is growing evidence that composite SF/AGN systems are common at medium to high redshift ($z$ $\ga$ 1--2; see e.g.\ \citealt{Dad2007, Del2013, Ree2016}).

The most reliable way to separate AGN cores (size $\ll$ 1\,kpc) from starbursts ($\gtrsim$ 1\,kpc) in intermediate redshift \mbox{($z$ $\sim$ 1)} galaxies is through deep ($\mu$Jy-level) imaging at milli-arcsec (mas) resolution using Very Long Baseline Interferometry (VLBI). 

The ECDFS is arguably the field in which the properties of sub-mJy RQ AGNs have been most extensively studied. Here we set constraints on the AGN contribution to the total radio emission in ECDFS sources classified as RQ AGNs using the Australian Long Baseline Array (LBA).

Throughout this letter, we adopt a standard flat $\Lambda$CDM cosmology with H$_0$ = 70\,km\,s$^{-1}$\,Mpc$^{-1}$ and $\Omega_M$ = 0.30.

\vspace{-0.2cm}
\section{Strategy} \label{sec:Strategy}

Our targets were selected from a 1.4\,GHz catalogue of 883 radio sources obtained from VLA observations of the ECDFS field \citep{Mil2013}, which has a typical rms sensitivity of $\sim$\,7.4\,$\mu$Jy and an angular resolution of 2.8\,$\times$\,1.6 arcsec$^2$. The availability of extensive and deep multi-band information from radio \citep{Huy2012, Fra2015} to X-ray band \citep{Leh2005, Xue2011} allowed the classification of all the sources from Miller and collaborators into three classes: RL AGNs, RQ AGNs, and SFGs \citep{Bon2013}.

The classification of \citet{Bon2013} is primarily based on the observed infrared-to-radio ratio \mbox{(\,$q_{24\,obs}$ = Log$(S_{24\,\text{$\mu$m}}/S_{1.4\,\text{GHz}})$\,)}. Any source that shows a significant radio excess with respect to an M82 template (redshifted up to $z$\,=\,10) is classified as RL. Sources with no significant radio excess are classified as RQ AGNs if they show clear evidence for AGN activity in the X-rays \mbox{($L_{2-8\,\text{keV}}$ > 10$^{42}$ erg\,s$^{-1}$)} or in the mid-infrared bands (i.e. the source lies inside the AGN wedge of the IRAC colour-colour diagram; \citealt{Don2012}). A quality flag (QF) was also defined to express the reliability of the classification from secure (QF = 3) to reasonable \mbox{(QF = 2)} and tentative \mbox{(QF = 1)} \citep{Bon2013}.

The ECDFS field has been observed with the VLBA by \citet{Mid2011}. However, the position of the field (Dec\,$\sim$\,$-28^\circ$) is not well suited for the VLBA, and the average elevation during the experiment was only $\sim$\,20$^\circ$. This limited the sensitivity of the observation: only sources brighter than 400--1000\,$\mu$Jy (depending on the position in the field) could be detected. The VLBA detected one RQ AGN with a flux density \mbox{> 400\,$\mu$Jy}, but the source was flagged only as reasonable (\mbox{QF = 2} in the catalogue of \citealt{Bon2013}).

\begin{figure}[!t]
\begin{minipage}{0.48\textwidth}
\centering
\includegraphics[scale=0.52]{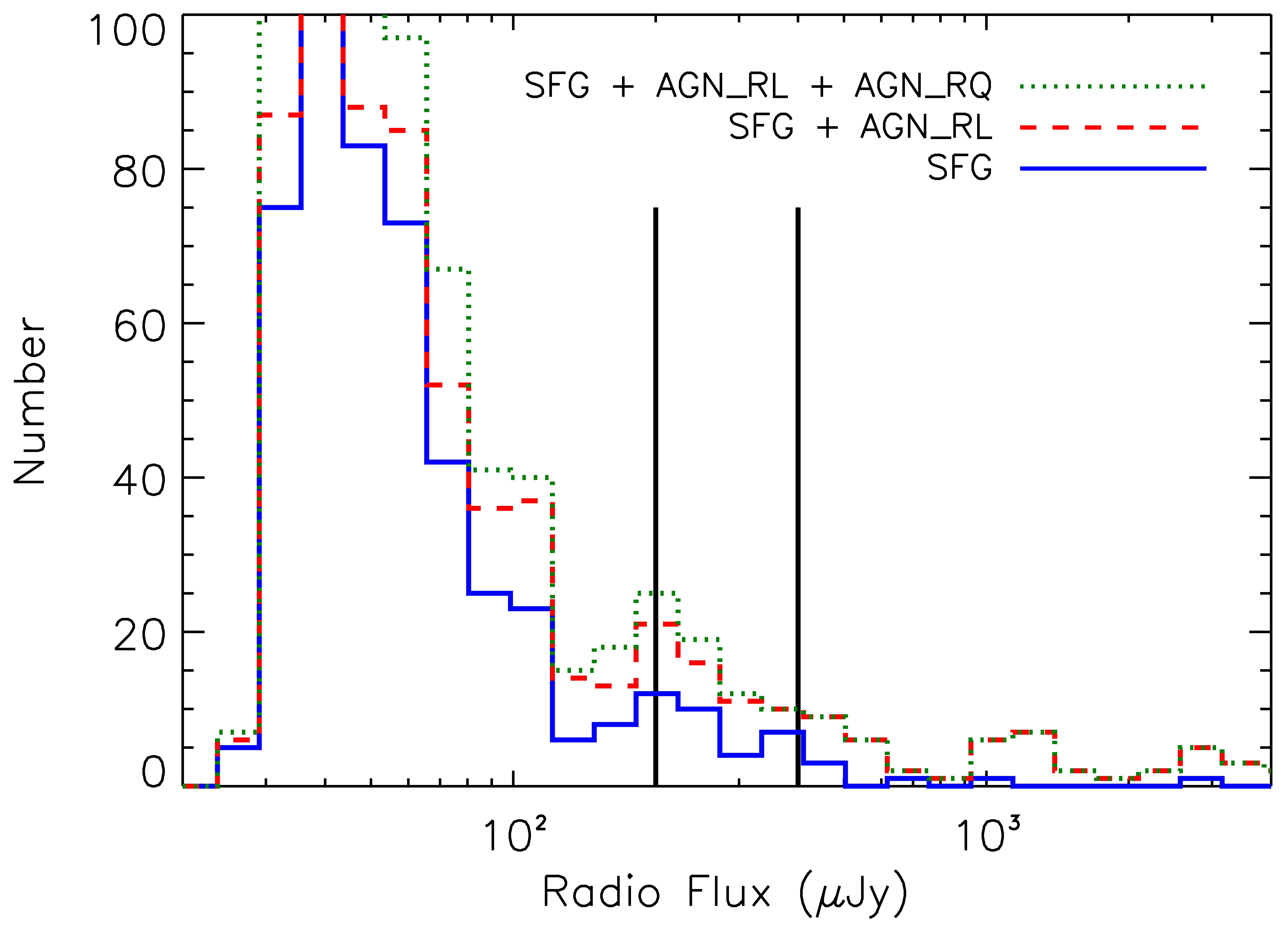}
\end{minipage}\hfill
\begin{minipage}{0.48\textwidth}
\centering
\caption{ \small{\label{fig:Bonzini_Distribution}
Stacked histogram of the radio flux density distributions of the sources in \citet{Bon2013}. The solid blue curve refers to SFGs, the red dashed curve adds to it the RL AGNs, the green dotted curve adds to them the secure (QF = 3) RQ AGNs. The two black vertical lines delimit the flux range of the sample we targeted (200--400\,$\mu$Jy).
} }
\end{minipage}
\end{figure}

Sources securely classified as RQ AGNs (QF = 3) are fainter than 400\,$\mu$Jy (see Fig.\,\ref{fig:Bonzini_Distribution}). Deeper VLBI observations are therefore needed to probe the RQ AGN component in the ECDFS field to determine whether AGN cores are present. Here we report VLBI observations with the Australian Long Baseline Array (LBA), which is better suited for observations of this southern field.

VLBI observations at $\mu$Jy level are challenging. Here we focus on four RQ AGNs with secure classification (QF = 3), selected from the brightest (with \mbox{200\,$\mu$Jy <  S$_{1.4\,GHz}$ < 400\,$\mu$Jy}) and unresolved (on arcsec-scale) sources, together with  a control sample of four RL AGNs selected to match the flux density and redshift distribution of the RQ AGNs. This control sample not only allows us to check our data analysis procedure, but also measures the fraction of RL AGN detected in LBA observations.

\begin{table}[!b]
\caption{\label{tab:Arrays} Observation runs of the project.}
\scriptsize
\centering
\renewcommand{\arraystretch}{1.2}
\begin{tabular}{c c c c c c c}
\hline\hline
Run &    Date    & t$_{Obs}$ & $\nu_{Obs}$ &   BW   &         Antennas           & Target(s) \\
    &            &   (hrs)   &    (GHz)    &  (MHz) &                            &           \\
(1) &    (2)     &    (3)    &     (4)     &   (5)  &           (6)              &    (7)    \\
\hline                                                                                                                                           
 A  & 09/03/2014 &   9.5     &   1.666     &   64   & AK, AT, Cd, Ho, Mp, Pa     &   RQ26    \\
    &            &           &             &        &                            &   RL106   \\
    &            &           &             &        &                            &   RL728   \\
 B  & 04/06/2014 &    11     &   1.650     &   32   &     AT, Cd,     Mp, Pa, Ti &   RQ174   \\
    &            &           &             &        &                            &   RQ851   \\
 C  & 26/11/2014 &    12     &   1.410     &   64   & AK, AT, Cd, Ho,     Pa, Ti &   RL183   \\
    &            &           &             &        &                            &   RL287   \\
 D  & 30/03/2015 &    10     &   1.410     &   64   &     AT, Cd, Ho, Mp, Pa     &   RQ851   \\
 E  & 31/03/2015 &     9     &   1.410     &   64   &     AT, Cd, Ho, Mp, Pa     &   RQ76    \\
\hline \\                           
\end{tabular}
\tablefoot{
\small{
t$_{Obs}$ = Observation time; $\nu_{Obs}$ = Average frequency of observation; BW = Total bandwidth; AK = ASKAP; AT = ATCA; Cd = Ceduna; Ho = Hobart; Mp = Mopra; Pa = Parkes; Ti = Tidbinbilla. Data from Run C were affected by  technical problems, and were not used.
}
}
\end{table}

\vspace{-0.2cm}
\section{Observations and data reduction} \label{sec:Observations_and_Data_Reduction}

The LBA observations took place in five different runs, between March 2014 and March 2015, for a total of 51.5 hours. Seven radio telescopes were requested for this experiment, with actual availability depending on the run (see Table\,\ref{tab:Arrays}). The data were correlated at the Pawsey Centre for Supercomputing.

To maximise sensitivity, we observed in cycles of about six minutes by switching between target and phase calibrator (QSO J0348-2749, about 3.85 degrees away). Every 1--1.5 hours, we observed the fringe finder (QSO B0208-512). When more than one target was observed, the six-minute cycles were shared between the targets, with the observing time inversely proportional to their flux density. Overall, $\sim$\,68\% of the time was devoted to the target(s), $\sim$\,25\% to the phase calibrator, and $\sim$\,7\% to the fringe finder.

\begin{figure*}[!ht]
\begin{minipage}{0.3\textwidth}
\centering
\includegraphics[width=140pt]{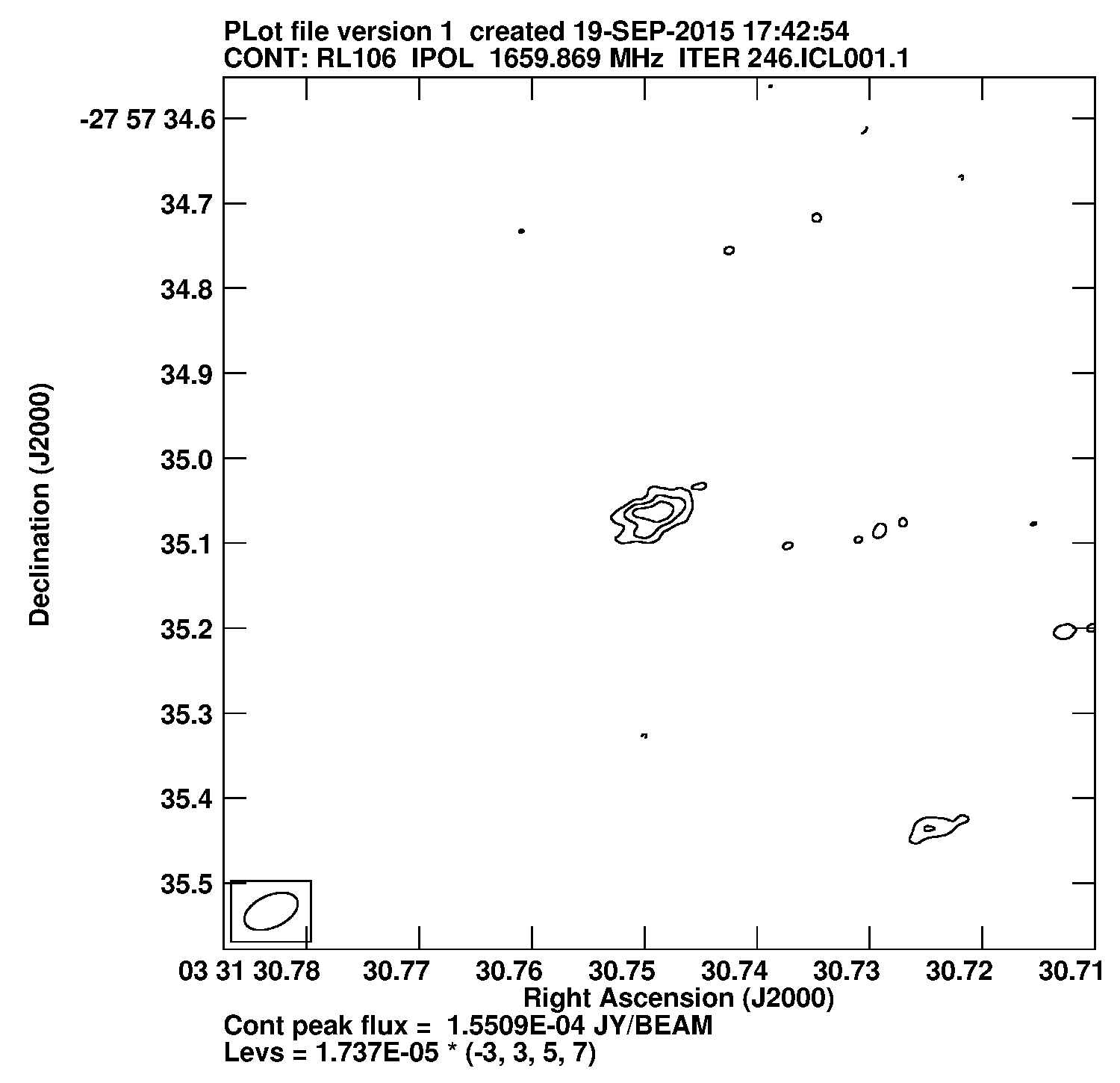}
\end{minipage}
\mbox{}
\begin{minipage}{0.05\textwidth}
\centering
\mbox{}
\end{minipage}
\mbox{}
\begin{minipage}{0.3\textwidth}
\centering
\includegraphics[width=140pt]{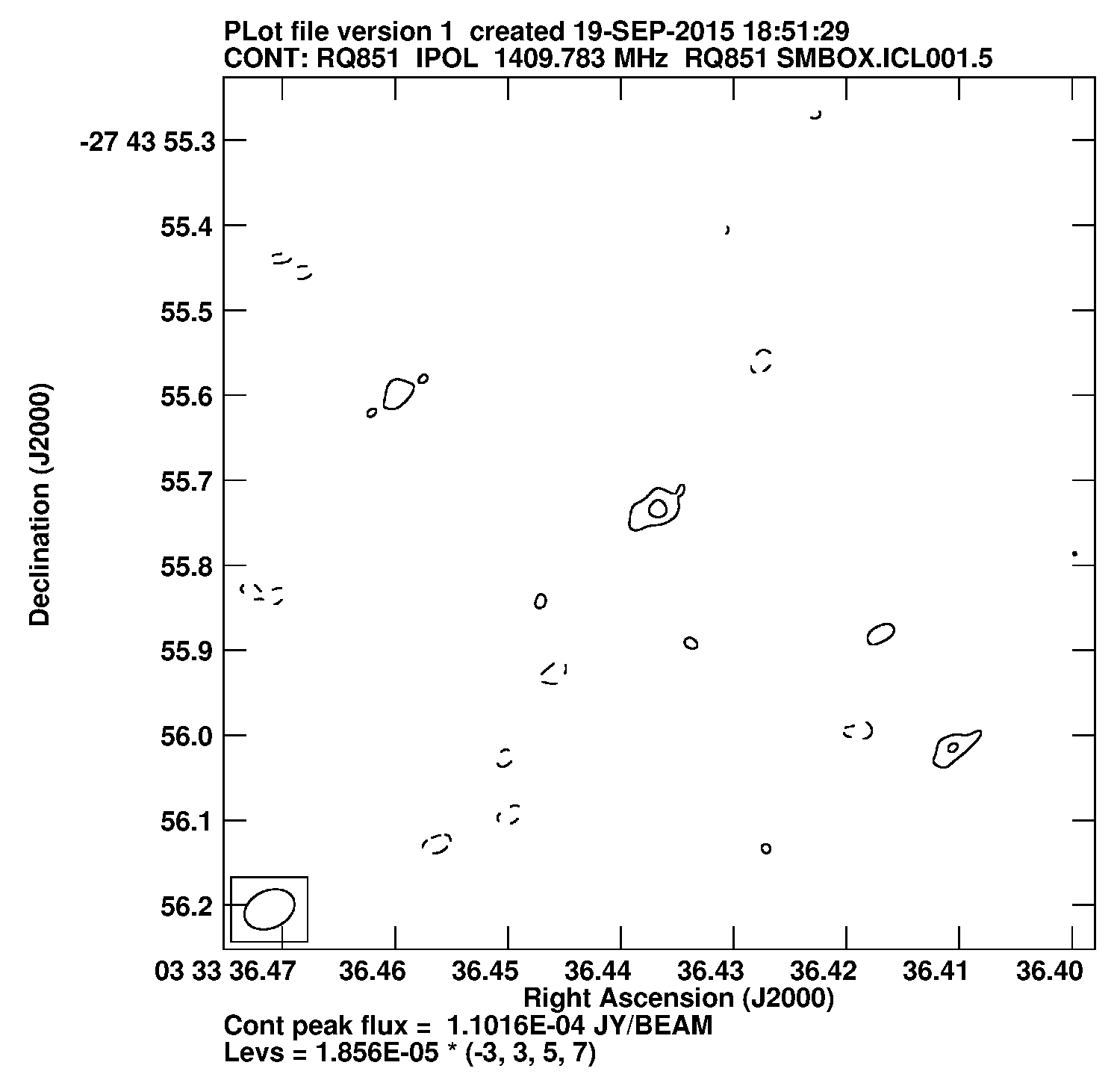}
\end{minipage}
\mbox{}
\begin{minipage}{0.3\textwidth}
\centering
\includegraphics[width=140pt]{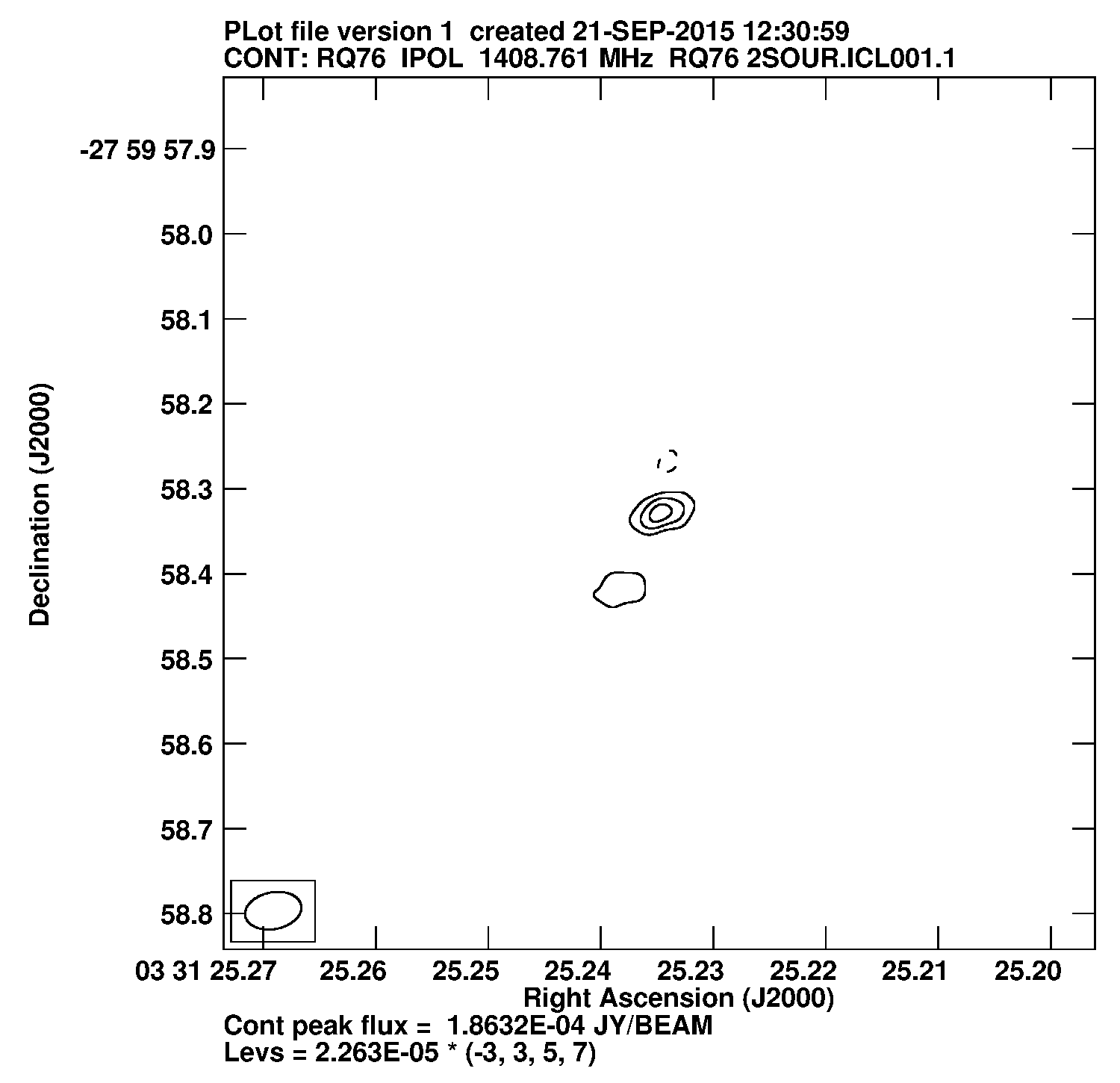}
\end{minipage} \\
\begin{minipage}{0.95\textwidth}
\centering
\caption{ \small{\label{fig:Detections}
Images of the VLBI-detected sources: RL\,106 ({\it Left}), RQ\,851 ({\it Centre}), and RQ\,76 ({\it Right}). All the isophotes are drawn at $-$3, 3, 5 and 7 times the image 1$\sigma$ noise (17.4, 18.6 and 22.6\,$\mu$Jy, respectively). All images are $\sim$\,1\,$\times$\,1 arcsec$^2$ wide.
} }
\end{minipage}
\end{figure*}

Calibration and imaging were carried out with the {\it Astronomical Image Processing System}\footnote{http://www.aips.nrao.edu/index.shtml}
(AIPS) software package. We first  applied onsource and manual flagging using the task \verb+UVFLG+. Ionospheric Faraday rotation and dispersive delay were corrected for using \verb+VLBATECR+, and cross-correlation amplitude spectra were calibrated from auto-correlation data using \verb+ACCOR+. We used \verb+CLCOR+ to correct the antenna gains and corrected for parallactic angle using the task \verb+VLBAPANG+. We then corrected the instrumental phases and delays using the fringe finder and \verb+VLBAMPCL+. Finally, we referred the phases of our targets to the phase of the phase calibrator through the task \verb+VLBAFRGP+. When needed, a band-pass calibration was applied with the task \verb+BPASS+. The a priori calibration was followed by a series of phase-only and amplitude+phase self-calibration loops on the phase calibrator, using the tasks \verb+IMAGR+ and \verb+CALIB+. This solution was then used to calibrate the multi-data file.

We imaged the UV data with the task \verb+IMAGR+ into 2048\,$\times$\,2048 pixel maps, with a pixel size of 1\,mas and natural weighting. The images were cleaned with the Cotton-Schwab algorithm in a circle of 512 pixels radius centred on the detections. Typical restoring beams are $\sim$65\,$\times$\,42\,mas$^2$.

The maximum angular scale to which our LBA observations are sensitive is \mbox{$\vartheta$ $\sim$ 0.33"} (A and B runs) and \mbox{$\vartheta$ $\sim$ 0.39"} (D and E runs). In the redshift range of our targets \mbox{(1.0 $\la$ $z$ $\la$ 2.9)}, this corresponds to a maximum linear scale of $\sim$\,2.7\,kpc (A and B runs) and $\sim$\,3.2\,kpc (D and E runs).

\vspace{-0.2cm}
\section{Results} \label{sec:Results}

Of the six sources successfully observed (two RL and four RQ AGNs), three were detected: one RL AGN (RL106) and two RQ AGNs (RQ851 and RQ76). The parameters of detected sources are listed in Table\,\ref{tab:Detections}, while a detailed descriptions of the targets is given below.

\vspace{-0.3cm}
\subsection{Detected sources} \label{subsec:Detected}

\hspace{12pt} {\it \textbf{RL106}} \ \ Source 106 of \citet{Mil2013} is classified as a RL AGN by \citet{Bon2013} because of its high radio excess (\,$q_{24\,obs}$ = $-$0.99$\pm$0.18; \citealt{Lut2011}) and its location in the IRAC colour-colour diagram, where it lies well within the region occupied by elliptical galaxies (\,Log($S_{5.8\,\text{$\mu$m}}/S_{3.6\,\text{$\mu$m}}$) = $-$0.48$\pm$0.04; \mbox{Log($S_{8.0\,\text{$\mu$m}}/S_{4.5\,\text{$\mu$m}}$) = $-$0.52$\pm$0.05}; \citealt{Lon2003, Sur2005}). We found a VLBI counterpart (Fig.\,\ref{fig:Detections}, left) coincident with the VLA position, with a peak flux density of 155\,$\mu$Jy (S/N ratio $\sim$\,9.1), or $\sim$43\% of the 360\,$\mu$Jy measured by the VLA \citep{Mil2013}. The missing radio flux presumably represents resolved AGN structures, such as jets.

{\it \textbf{RQ851}} \ \ Source 851 of \citet{Mil2013} is classified as a RQ AGN by \citet{Bon2013} because it has no radio excess (\,$q_{24\,obs}$ = 0.34$\pm$0.03; \citealt{Lon2003, Sur2005}) and is located in the the AGN wedge of the IRAC colour-colour diagram (\,Log($S_{5.8\,\text{$\mu$m}}/S_{3.6\,\text{$\mu$m}}$) = 0.26$\pm$0.02; \mbox{Log($S_{8.0\,\text{$\mu$m}}/S_{4.5\,\text{$\mu$m}}$) =} 0.14$\pm$0.01; \citealt{Lon2003, Sur2005}). We detected a VLBI counterpart (Fig.\,\ref{fig:Detections}, centre) $\sim$\,0.1" away from the VLA position, with a peak flux density of 110\,$\mu$Jy, or $\sim$50\% of the 222\,$\mu$Jy measured by the VLA \citep{Mil2013}. Our VLBI detection has a S/N ratio $\sim$\,5.5, and we therefore consider this detection as tentative.

{\it \textbf{RQ76}} \ \ Source 76 of \citet{Mil2013} is classified as a RQ AGN by \citet{Bon2013} because it has no radio excess (\,\mbox{$q_{24\,obs}$ = 0.53$\pm$0.03}; \citealt{Lon2003, Sur2005}), a powerful X-ray luminosity (\,\mbox{$L_{2-8\,\text{keV}}$ = $(4.2\pm0.3)\,\times\,10^{44}$\,erg\,s$^{-1}$}; \citealt{Leh2005}), and is located in the AGN wedge of the IRAC colour-colour diagram (\,Log($S_{5.8\,\text{$\mu$m}}/S_{3.6\,\text{$\mu$m}}$) = 0.50$\pm$0.03; Log($S_{8.0\,\text{$\mu$m}}/S_{4.5\,\text{$\mu$m}}$) = 0.60$\pm$0.02; \citealt{Lon2003, Sur2005}). We detected a VLBI counterpart (Fig.\,\ref{fig:Detections}, right) $\sim$\,1.2" away from the VLA position, with a peak flux density of 186\,$\mu$Jy (S/N ratio $\sim$\,8.1), or $\sim$69\% of the 269\,$\mu$Jy of the VLA flux density \citep{Mil2013}. The offset between the two radio positions is significant, but our detection lies inside the (large) positional error ellipse of the radio source (2.9\,$\times$\,1.6\,arcsec$^2$) reported by \citet{Mil2013}, is only 0.77" from the HST counterpart \citep{Rix2004}, and there is no other radio emission detected by the VLA, which is consistent with the measured VLBI detection. We therefore conclude that the VLBI detection is unambiguously the same source as the VLA detection, and the offset may be caused by extended emission, such as an AGN jet.

\vspace{-0.3cm}
\subsection{Undetected sources} \label{subsec:Undetected}

\hspace{12pt} {\it \textbf{RQ26}} \ \ Source 26 of \citet{Mil2013} is classified as a RQ AGN by \citet{Bon2013} because of its robust IR excess and location in the AGN wedge of the IRAC colour-colour diagram. We found no VLBI counterpart down to a 3$\sigma$ flux density level of $\sim$\,157\,$\mu$Jy, or $\sim$\,50\% of the VLA flux density of 318\,$\mu$Jy \citep{Mil2013}.
                                         
{\it \textbf{RL728}} \ \ Source 728 of \citet{Mil2013} is classified as a RL AGN by \citet{Bon2013} because of a slight radio excess, strong X-ray emission, and location in the the AGN wedge of the IRAC colour-colour diagram. We found no VLBI counterpart down to a 3$\sigma$ flux density level of $\sim$109\,$\mu$Jy, or $\sim$\,30\% of the VLA flux density  of 326\,$\mu$Jy \citep{Mil2013}. This source is discussed in detail in Sect.\,\ref{sec:Discussion}.

{\it \textbf{RQ174}} \ Source 174 of \citet{Mil2013} is classified as a RQ AGN by \citet{Bon2013} because of a significant IR excess and high X-ray luminosity. We found no VLBI counterpart down to a 3$\sigma$ flux density level of $\sim$\,125\,$\mu$Jy, or $\sim$\,42\% of the VLA flux density of 300\,$\mu$Jy \citep{Mil2013}.

\begin{table*}[!t]
\caption{\label{tab:Detections} Summarised characteristics of our detections.}
\scriptsize
\centering
\renewcommand{\arraystretch}{1.2}
\begin{tabular}{c c c c c c c c c c | c c}
\hline\hline
 Target &  $S_{VLBI}$  & $S_{VLBI}/S_{VLA} $ &     r.m.s.     & $K$-corr.\tablefootmark{a} $L_{1.4\,GHz}$  &   Restoring beam    &         T$_B$        &          $z$          &       Linear scale      &         Host          &   $\nu^{YSN}$   &   $\nu^{SNR}$   \\
        &   ($\mu$Jy)  &                     & ($\mu$Jy/beam) &         ($\times 10^{23}$\,W/Hz)           &      (mas$^2$)      & ($\times 10^{4}$\,K) &                       &           (pc)          & type\tablefootmark{e} & (SN\,yr$^{-1}$) & (SN\,yr$^{-1}$) \\
   (1)  &      (2)     &         (3)         &      (4)       &                    (5)                     &         (6)         &          (7)         &          (8)          &            (9)          &         (10)          &      (11)       &      (12)       \\
\hline                                                                                                                                                                                                                                                               
 RQ26   &  $\la$ 157   &     $\la$ 0.49      &       52       &                    ...                     &         ...         &          ...         & 1.59\tablefootmark{b} &           ...           &          --           &      ...        &      ...        \\
 RL106  & 155 $\pm$ 29 &  0.43 $\pm$ 0.08    &       17       &                 6.6\,$\pm$\,1.2            & $\sim 67 \times 38$ &    2.7\,$\pm$\,0.5   & 1.06\tablefootmark{c} &  $\la$ 544 $\times$ 308 &          P            &   $\sim$\,126   &   $\sim$\,11    \\
 RL728  &  $\la$ 109   &     $\la$ 0.33      &       36       &                    ...                     &         ...         &          ...         & 1.08\tablefootmark{b} &           ...           &          SB           &      ...        &      ...        \\
 RQ174  &  $\la$ 125   &     $\la$ 0.42      &       42       &                    ...                     &         ...         &          ...         & 2.85\tablefootmark{c} &           ...           &          MS           &      ...        &      ...        \\
 RQ851  & 110 $\pm$ 26 &  0.50 $\pm$ 0.12    &       19       &                 9.7\,$\pm$\,2.3            & $\sim 62 \times 44$ &    2.5\,$\pm$\,0.6   & 1.35\tablefootmark{d} &  $\la$ 521 $\times$ 370 &          MS           &   $\sim$\,163   &   $\sim$\,15    \\
 RQ76   & 186 $\pm$ 36 &  0.69 $\pm$ 0.14    &       23       &                 17.2\,$\pm$\,3.3           & $\sim 67 \times 43$ &    4.0\,$\pm$\,0.8   & 1.38\tablefootmark{b} &  $\la$ 564 $\times$ 362 &          SB           &   $\sim$\,290   &   $\sim$\,26    \\
\hline
\end{tabular}
\tablefoot{
\tablefoottext{a}{Rest-frame 1.4\,GHz luminosities were obtained by using measured 1.4-5\,GHz spectral indices, based on data from \citet{Hal2014} and \citet{Huy2012}. For RQ851 and RQ76, 5\,GHz information is lacking, and we adopted the value of $-$0.75 (e.g., \citealt{Kuk1998});}
\tablefoottext{b}{spectroscopic, \citet{Sil2010};}
\tablefoottext{c}{photometric, \citet{Raf2011};}
\tablefoottext{d}{photometric, \citet{Tay2009};}
\tablefoottext{e}{from \citet{Bon2015}: P=Passive; SB= starbust galaxy; MS= main sequence galaxy.}
}
\end{table*}

\vspace{-0.2cm}
\section{Discussion and conclusions} \label{sec:Discussion}

VLBI detections are usually considered as unambiguous evidence of AGN activity \citep{Kew2000}. However, sensitivity improvements mean that VLBI can now detect a surface brightness of T$_B$\,<\,10$^5$\,K, which can in principle have another origin. We therefore examined here whether this might be caused by compact H\,II regions, radio supernovae, or compact supernova remnants.

Compact H\,II regions are characterised at radio wavelengths by thermal free-free emission with \mbox{$T_B \sim 10^{4-5}$\,K} and sizes of a few pc, resulting in a radio luminosity < 10$^{20}$\,W\,Hz$^{-1}$ \citep{Hug2007}, far lower than the luminosities of our VLBI detected sources (10$^{23-25}$\,W\,Hz$^{-1}$). To rule out the possibility that we detected bright supernova remnants or radio supernovae, we estimated the supernova rate corresponding to our radio luminosities. Following \citet{Kew2000}, we estimated the rate of radio-young supernovae ($\nu^{YSN}$) and of old supernova remnants ($\nu^{SNR}$) and list the result in Table\,\ref{tab:Detections} (Cols.\ 11 and 12). These values greatly exceed the rates for even the most extreme starburst galaxies (\,$\lesssim$\,0.4\,SN\,yr$^{-1}$; \citealt{Man2003}). We are therefore confident that the emission we detect in our sources is due to AGN activity.

Recently, \citet{Bon2015} revisited the FIR SFR estimates of the ECDFS sample using the \textit{Herschel} PACS data. This allowed them to obtain more reliable estimates of the source radio excess than those that were based on the $q_{24}$ parameter alone \citep{Bon2013}.

All the RQ AGNs we targeted lie below the radio/FIR correlation derived by \citet{Bon2015}, presumably due to our bright flux limit ($S_{VLA} \gtrsim$ 0.2\,mJy). On the other hand, the two LBA-detected RQ AGNs show radio excesses that are fully consistent with the overall distribution of the RQ AGN sub-sample around the correlation (i.e. they are not outliers). This may indicate that compact radio cores are a common component of RQ AGNs, at least for those lying below the radio/FIR correlation. \citet{Bon2015} discussed the larger dispersion of RQ AGNs around the radio/FIR correlation with respect to the dispersion of SFGs and hypothesised that it might be due to contamination by AGN emission. Here we provide direct evidence that this is the case. When radio cores are removed, the two detected RQ AGNs fall nicely on the radio/FIR correlation. It is also interesting to note that our undetected RL AGN (RL728) is associated with an SB galaxy and lies close to the radio/FIR correlation, based on the new FIR SFR estimation of \citet{Bon2015}. We can therefore argue that this is most likely a misclassified source.

To summarise, we found direct evidence that some of the AGNs classified as radio-quiet in deep radio samples have mas-scale, AGN-triggered radio cores, at least at flux densities $S_{VLA} \gtrsim$ 0.2\,mJy. Such cores can account for a significant fraction of the total radio emission (up to $\sim$70\% in RQ76). Although limited in a statistical sense, our findings are supported by similar results by \citet{Her2016}, who found VLBI cores in RQ QSOs in the COSMOS field. Further support comes from the 12 RQQ sources (according to the classification of \citealt{Bon2013}) observed by \citet{Chi2013}, four of which were detected with VLBI.


On the other hand, it is clear that at redshifts $z$ $\ga$ 1 many RL AGNs are hosted by galaxies with strong star formation \citep{Ale2005, Dad2007, Nor2007, Ale2008} and that these star-forming RL AGNs dominate the most radio-luminous AGNs at high redshift \citep{Ree2016}.

We therefore expect that the focus of future studies will shift from determining the nature of RQ sources to measuring the fraction of bolometric luminosity that is contributed by AGN and SF activities. However, further work is needed to understand the factors that determine whether a RQ AGN has a VLBI-detectable core.

\nocite{Wri2006}

\vspace{0.2cm}

\begin{acknowledgements}
The authors thanks the anonymous referee for the valuable comments, that allowed us to improve the discussion of our results.

AM is responsible for the content of this publication.

AM acknowledges funding by a Cotutelle International Macquarie University Research Excellence Scholarship (iMQRES).

IP and RPN acknowledge support from the Ministry of Foreign Affairs and International Cooperation, Directorate General for the Country Promotion (Bilateral Grant Agreement ZA14GR02 - Mapping the Universe on the Pathway to SKA).

The Australian SKA Pathfinder is part of the Australia Telescope National Facility which is managed by CSIRO. Operation of ASKAP is funded by the Australian Government with support from the National Collaborative Research Infrastructure Strategy. Establishment of the Murchison Radio-astronomy Observatory was funded by the Australian Government and the Government of Western Australia. ASKAP uses advanced supercomputing resources at the Pawsey Supercomputing Centre. We acknowledge the Wajarri Yamatji people as the traditional owners of the Observatory site

The Australia Telescope Compact Array (ATCA), the Parkes radio telescope, the Mopra radio telescope, and the Long Baseline Array, are part of the Australia Telescope National Facility which is funded by the Australian Government for operation as a National Facility managed by CSIRO.

This work was supported by resources provided by the Pawsey Supercomputing Centre with funding from the Australian Government and the Government of Western Australia.

This research has made use of NASA's Astrophysics Data System.
\end{acknowledgements}

\vspace{-0.8cm}
    \bibliographystyle{aa} 
    \bibliography{Biblio} 

\end{document}